\documentclass[reprint,amsmath,amssymb,aps]{revtex4-1}

\usepackage{graphicx}
\usepackage{dcolumn}
\usepackage{bm}

\begin{document}


\title{Beyond ideal two-dimensional metals: Edges, vacancies, and polarizabilities}

\author{Janne Nevalaita}
\author{Pekka Koskinen}%
 \email{pekka.koskinen@iki.fi}
\affiliation{%
Department of Physics, Nanoscience Center, University of Jyv{\"a}skyl{\"a}, 40014 Jyv{\"a}skyl{\"a}, Finland
}%

\date{\today}

\begin{abstract}

Recent experimental discoveries of graphene-stabilized patches of two-dimensional (2D) metals have motivated also their computational studies. However, so far the studies have been restricted to ideal and infinite 2D metallic monolayers, which is insufficient because in reality the properties of such metallic patches are governed by microstructures pervaded by edges, defects, and several types of perturbations. Here we use density-functional theory to calculate edge and vacancy formation energies of hexagonal and square lattices of 45 elemental 2D metals. We find that the edge and vacancy formation energies are strongly correlated and decrease with increasing Wigner-Seitz radii, analogously to surface energies. Despite a radical reduction in atomic coordination numbers, the 2D and 3D vacancy formation energies and work functions are nearly the same for each metal. Finally, static polarizabilities reveal a clear cubic dependence on bond length. These trends provide useful insights when moving towards reality with elemental 2D metals.


\end{abstract}

\maketitle



\section{Introduction}

Since the discovery and success of graphene~\cite{novoselov04,novoselov05,zhang05, geim07,novoselov07,bolotin08,balandin08,lee08,geim09,ferrari15,franklin15}, much research has been devoted to finding new two-dimensional (2D) systems~\cite{butler13,bhimanapati15,lin16, zeng18}. The family of 2D materials has increased to include, for example, halides and transition metal chalcogenides~\cite{miro14}. Many of the known 2D structures consist of tightly bound monolayers held together by van der Waals forces and are therefore relatively easy to isolate by exfoliation~\cite{coleman11,nicolosi13}. However, synthesizing 2D materials from non-layered bulk structures requires different approaches~\cite{ma18}. Recent experiments have found 2D structures having non-layered 3D bulk counterparts, such as free-standing 2D iron patches grown inside graphene nanopores~\cite{zhao14}. The existence of 2D iron is surprising considering the metallic bonding in bulk iron that lacks the strongly directional character of covalent bonding, usually associated to 2D materials. In addition to being interesting for basic research, metallic 2D materials have several potential applications~\cite{ling15,fan15,chen18}, including catalysis and gas-sensing~\cite{pan14,deng16}.

The experimental evidence for 2D structures composed of metal atoms has motivated much computational research. Multiple elements and 2D lattices have been studied including Au, Ag and Cu monolayers~\cite{yang15b,yang15,yang16}, transition metal monolayers~\cite{hwang18}, and our recent study of elemental monolayers from 45 metals in three 2D lattices~\cite{nevalaita18}. However, so far most of the computationally studied free-standing elemental monolayers have been ideal and periodic in the plane of the atoms. These calculations are relevant for sufficiently large, ordered systems. Yet under realistic conditions 2D systems will have defects, such as edges and vacancies~\cite{zou15}. While the edges can be stabilized by supporting materials, edge formation energies provide information about the stabilities of finite systems compared to the periodic ones. While the vacancy formation energy is related to the stability of a 2D structure, it is also connected to the atom mobility and is useful in identifying promising elements for 2D liquids~\cite{koskinen15,yang16b}. Realistic systems can also have perpendicular perturbations. For example, 2D systems can host adsorbates~\cite{antikainen17}, can be grown on surfaces~\cite{yin09,zhang10,li13,yin15a,yin15b,zhu15,duan16,kochat18}, or can be a part of a layered heterostructure~\cite{li16}. In these cases the interactions perpendicular to the monolayers changes the properties of the ideal free-standing 2D lattices.

\begin{figure}
\includegraphics[width = 8.6 cm]{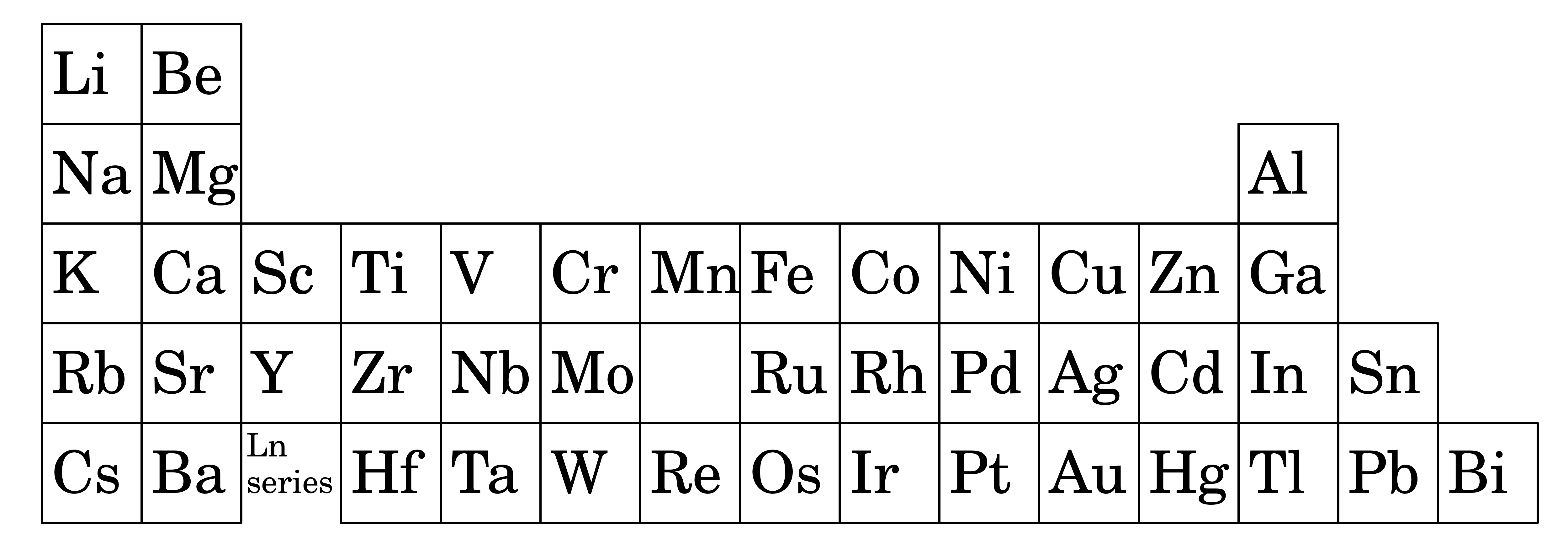}
\caption{Part of the periodic table with the chemical symbols of the 45 studied metals.}
\label{fig:metals}
\end{figure}

\begin{figure}
\includegraphics[width = 8.6 cm]{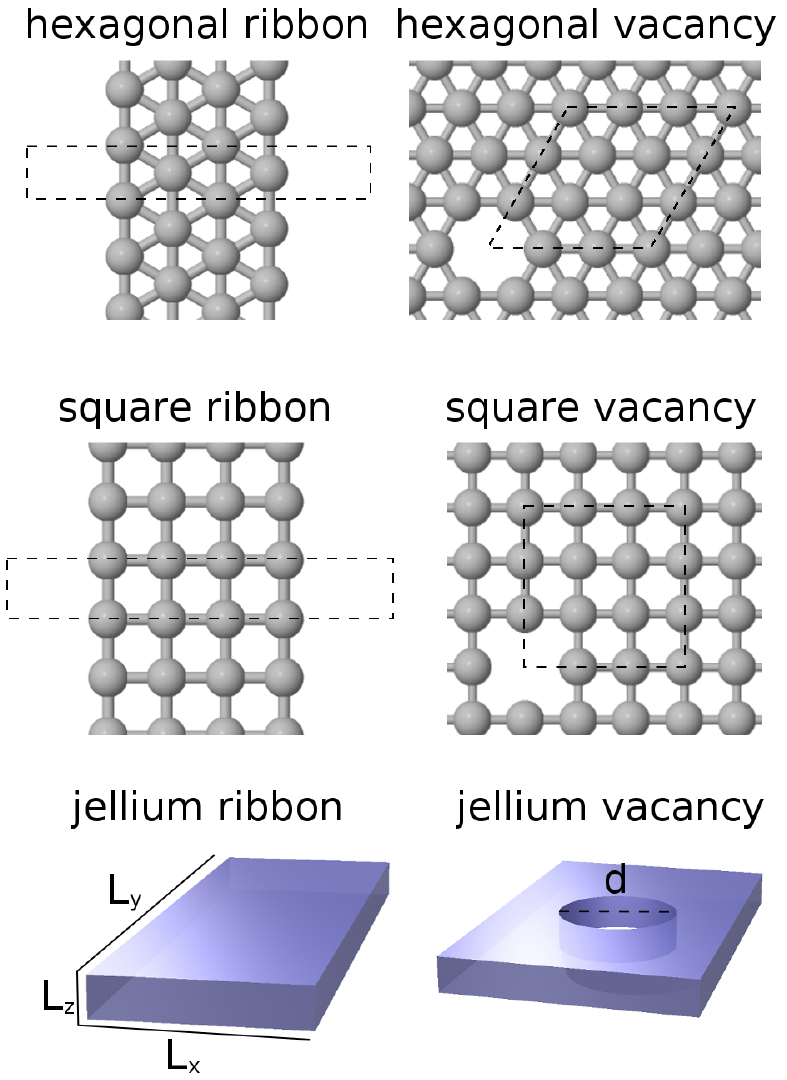}
\caption{Sketches of calculated structures. The dashed line indicates the varying computational cell.}
\label{fig:sketch}
\end{figure}

In this paper we aim to address how the finite size, vacancies, and simple perpendicular perturbations affect the properties of ideal free-standing monolayers of 2D metals. Using a density-functional theory (DFT) approach, we calculate the edge and vacancy formation energies for 45 metals (\mbox{Fig. \ref{fig:metals}}) in hexagonal and square geometries (\mbox{Fig. \ref{fig:sketch}}). We correlate these properties with the ones of conventional 3D bulk structures and find that the edge formation energies are related to the surface energies, both decreasing with increasing Wigner-Seitz radii. Despite the drastic changes in coordination numbers, the 2D vacancy formation energies for many metals are close to 3D vacancy formation energies. Further, since vacancy can be considered to consist of a round edge encircling the missing atom, the vacancy formation energies are related to the edge energies. As a measure of sensitivity to perpendicular perturbations, we also consider the static polarizability of 2D monolayers and find that the polarizability per atom increases with increasing 3D bond length. The work functions of hexagonal 2D films are relatively close to those measured for polycrystalline 3D samples.

\section{Computational methods}

The edge and vacancy formation energies were obtained from total energies calculated with the density-functional approach as implemented in the GPAW-code~\cite{mortensen05,enkovaara10}. For consistency with respect to earlier work, the exchange and correlation energies were approximated with the Perdew-Burke-Ernzerhof (PBE) functional~\cite{perdew96}. Also previously converged computational parameters and lattice constants were used~\cite{nevalaita18}. All structures were calculated without relaxation using ideal bond lengths. The plane-wave cut-off was \mbox{800 eV} and \mbox{5 \AA} vacuum region separated atoms from the non-periodic unit-cell edges. The atomic ribbons were modeled with 1$\times$12$\times$1 Monkhorst-Pack k-point sampling~\cite{monkhorst76,pack77} and vacancy formation energies were calculated with a constant k-point density in the atomic plane. The polarizabilities of monolayers were calculated with dipole-layer corrections. Jellium calculations were done as spin-compensated and rest as spin-polarized.

\section{Results}

\subsection{Edge energies decrease with increasing Wigner-Seitz radii}

\begin{figure}
\includegraphics[width = 8.6 cm]{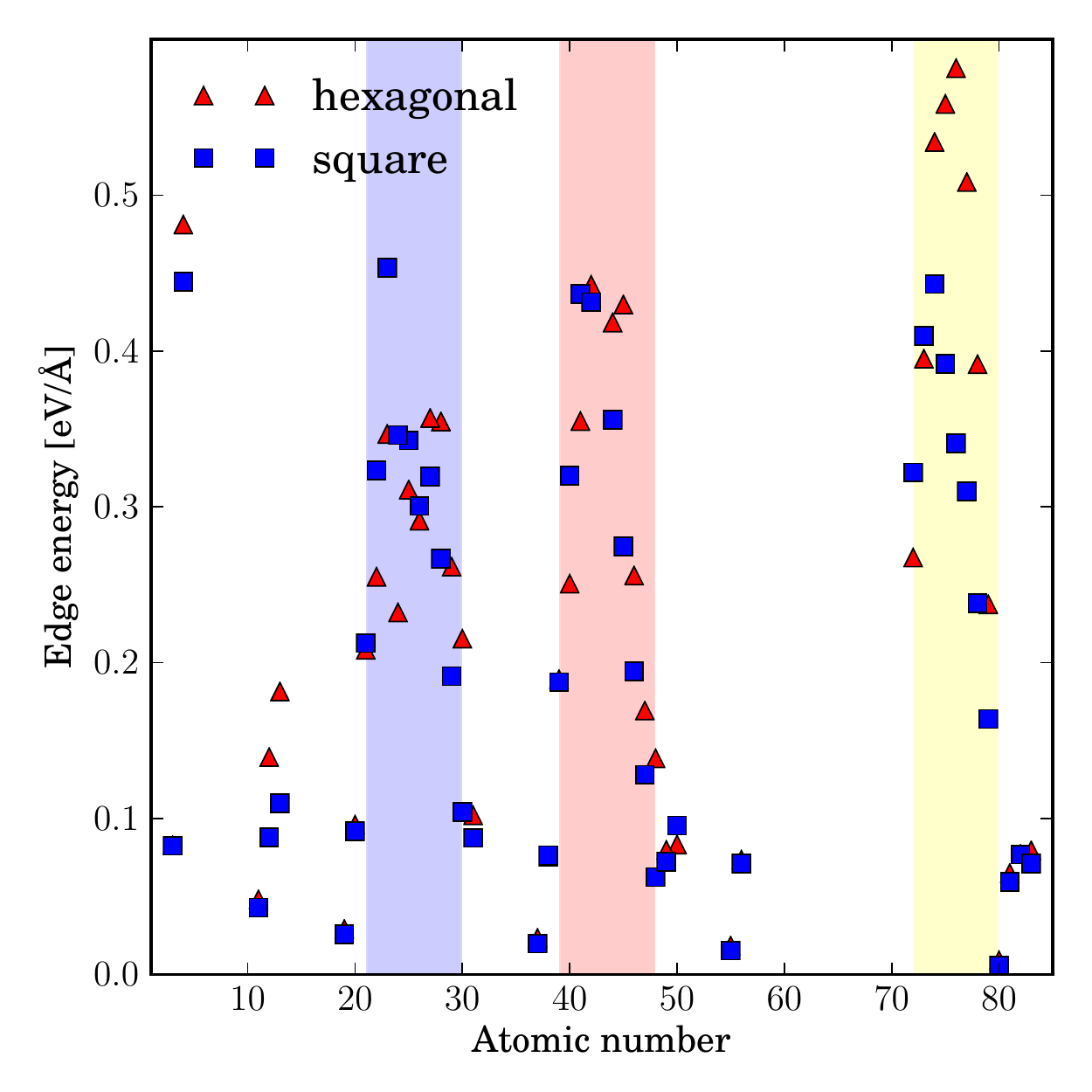}
\caption{Edge energies for hexagonal and square ribbons as a function of atomic number. Regions shaded with blue, red, and yellow correspond to 3d-, 4d-, and 5d-series, respectively.}
\label{fig:E_edge}
\end{figure}

We begin by considering the edge energy of all 45 metals in hexagonal and square geometries. Spin polarization is taken to account but since most systems are nonmagnetic we focus on other properties, starting with the edge energy~\cite{koskinen08} defined as 

\begin{equation}
\epsilon_{\mathrm{edge}} = \lim_{N\to\infty}\frac{E_r(N)-N\epsilon_{\mathrm{2D}}}{L_{\mathrm{edge}}},
\label{eq:edgelim}
\end{equation}
where $\epsilon_{\mathrm{2D}}$ is the energy per atom in periodic 2D structure, $E_r(N)$ the energy of a ribbon with $N$ atoms and $L_{\mathrm{edge}}$ the total length of the edge i.e. twice the length of the computational cell in the periodic direction. To get the edge energy one could choose a ribbon of specific width, calculate its energy, and subtract the corresponding 2D lattice energy. In this case the result would depend on the width of the chosen ribbon. Fortunately, this dependence can be removed with the ansatz

\begin{equation}
E_r(N) = N\epsilon_{\mathrm{2D}}+L_{\mathrm{edge}}\epsilon_{\mathrm{edge}}
\label{eq:edge}
\end{equation}
from which the edge energy is obtained by calculating ribbons of varying widths and fitting the properties $\epsilon_{\mathrm{edge}}$ and $\epsilon_{\mathrm{2D}}$ simultaneously. Further, comparing the fitted 2D lattice energy $\epsilon_{\mathrm{2D}}$ to one from a periodic calculation gives a convergence test for the edge energy. This method is a 2D analog of a similar approach for determining surface energies~\cite{fiorentini96}. The energy of a ribbon as a function of its width is linear already for very narrow ribbons, indicating that the ansatz~\eqref{eq:edge} holds already for small $N$. To rationalize this ansatz we consider a simple model of non-interacting electrons in a 1D box. To model a ribbon with varying width $L_x$ we set the number of particles $N$ proportional to the width of the well $N=\lambda L_x$. The energy as a function of $N$ in atomic units is

\begin{equation}
E_{\mathrm{box}}(N)=\sum^{N/2}_{n=1}\frac{\pi^2n^2}{L_x^2}=\pi^2\lambda^2\left(\frac{N}{24}+\frac{1}{8}+\frac{1}{12N}\right),
\end{equation}
which displays the observed linear behavior for large $N$. Further, second term on the rightmost side is independent of $N$ and corresponds to the edge energy. We expect that similar calculation with a 3D box introduces some complications, but leaves the general trend unaffected~\cite{wu08}.

The edge energies obtained this way are in general high near the middle of the d-series and particularly high for 5d-metals in hexagonal structures (\mbox{Fig. \ref{fig:E_edge}}). As discussed in our previous work~\cite{nevalaita18}, metals near the middle of the d-series have occupied bonding orbitals and unoccupied antibonding orbitals~\cite{hoffmann88}. This makes their bonds stronger and edge energies higher. The trend is qualitatively similar for both hexagonal and square lattices. The previously reported value of \mbox{0.2 eV/\AA} for Au agrees with our result~\cite{antikainen17}.

\begin{figure}
\includegraphics[width = 8.6 cm]{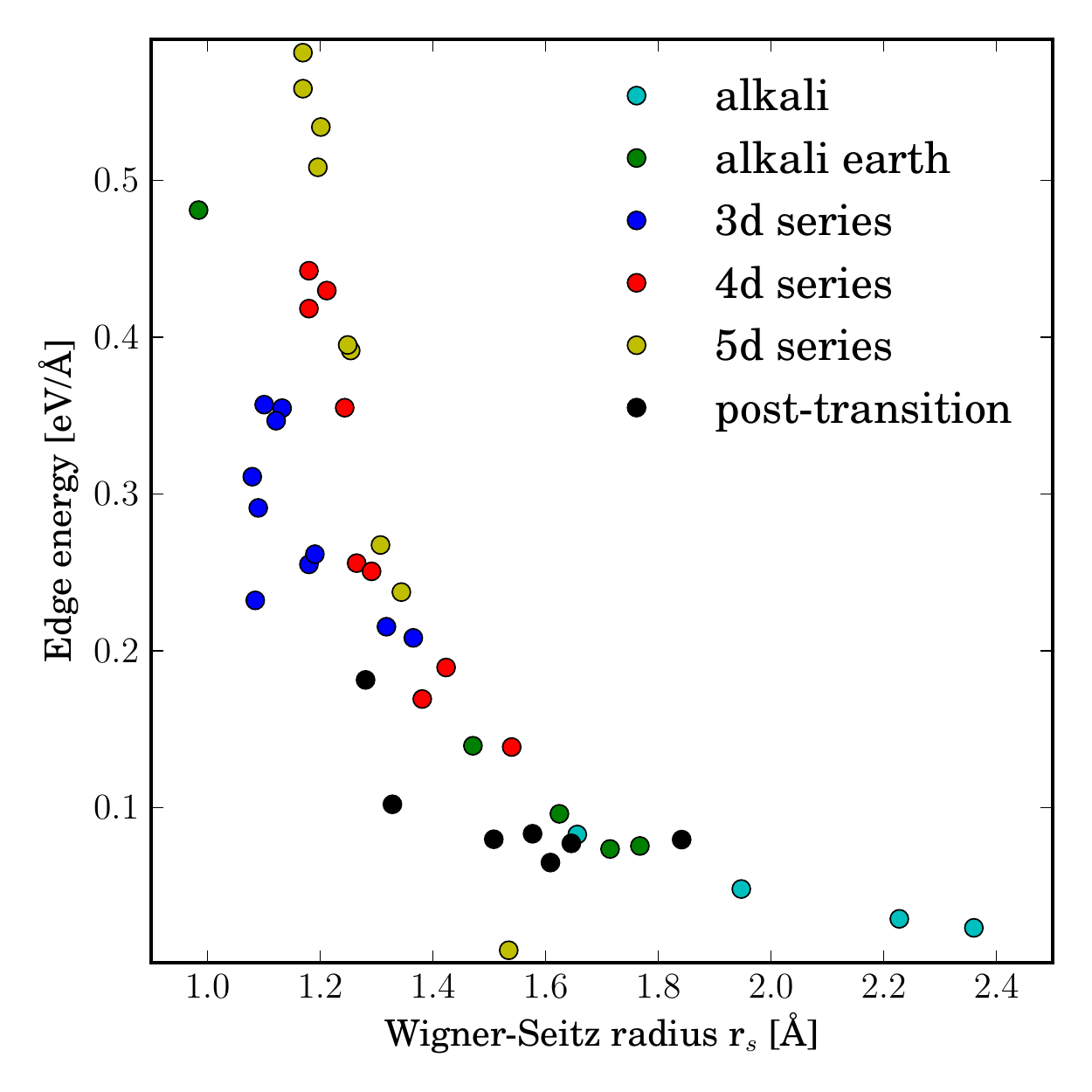}
\caption{Edge energies of hexagonal lattice as a function of $r_s$. Light blue, green, blue, red, yellow, and black correspond to alkali, alkali earth, 3d, 4d, 5d and post-transition metals, respectively. The $r_s$ values are from ref~\cite{perrot94}.}
\label{fig:rs_edge}
\end{figure}

Next, we consider edge energies as a function of the Wigner-Seitz radius $r_s$ defined by the equation

\begin{equation}
\frac{V}{N}=\frac{4\pi r_s^3}{3},
\label{eq:rs}
\end{equation}
where $N$ is the number of valence electrons in volume $V$. This new viewpoint emphasizes how the edge energies span almost two orders of magnitude and display roughly monotonic decrease with increasing $r_s$ (\mbox{Fig. \ref{fig:rs_edge}}). The trend holds especially well for simple metals. Most important, similar behavior has been reported for surface energies~\cite{perdew90}. We conclude that metals with high surface energies will have high edge formation energies, a trend that calls for closer inspection.

\begin{figure}
\includegraphics[width=8.6 cm]{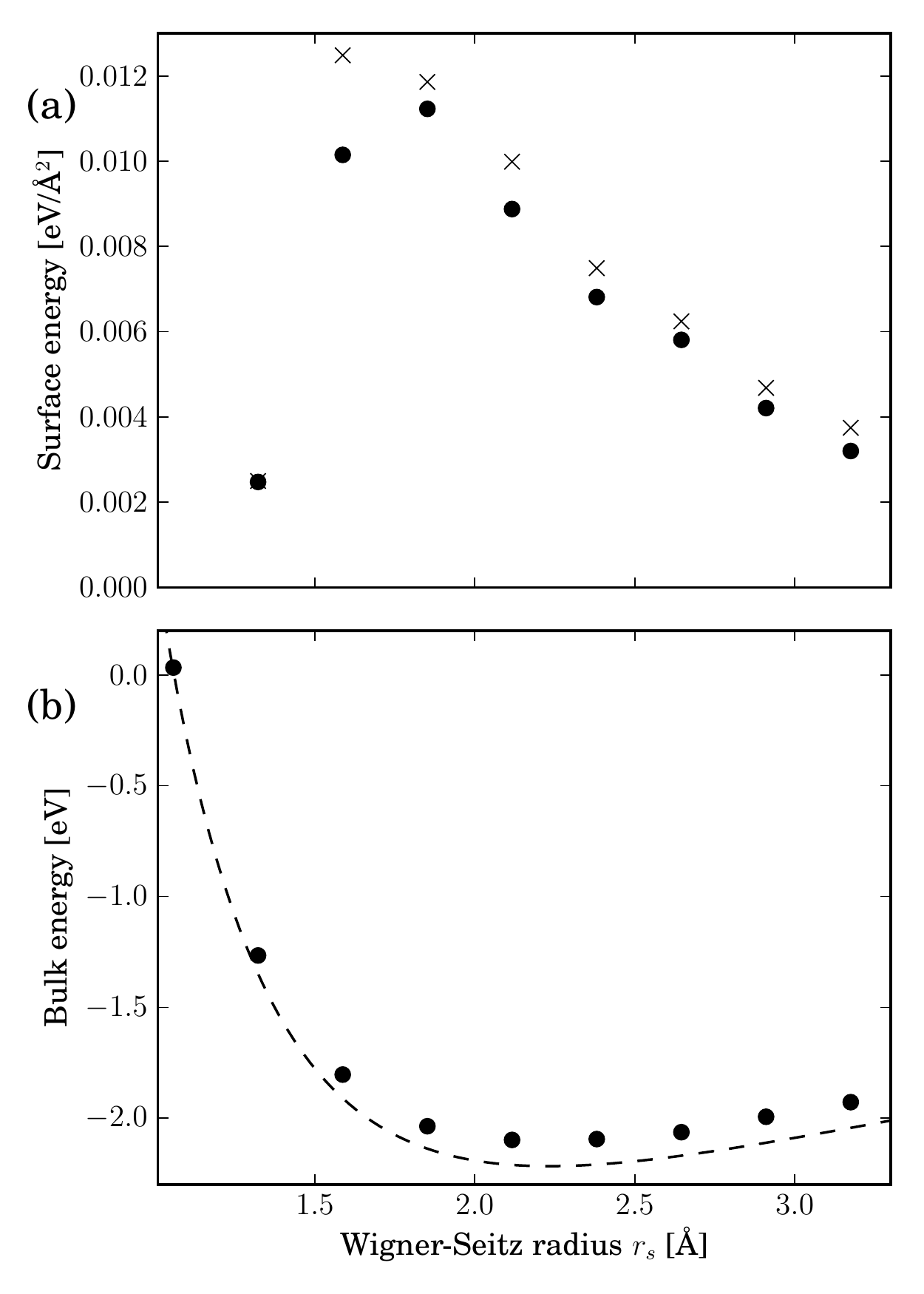}
\caption{a) Calculated surface energy from fit to equation~\eqref{eq:ribbon} (circles) and surface energies calculated by Lang and Kohn~\cite{lang70} (crosses).Values for $r_s=1.1$ \AA \ (-0.06 eV/\AA$^2$ by Lang and Kohn, -0.05 eV/\AA$^2$ by us) are omitted for easier visibility. b) Calculated bulk energy  obtained from equation~\eqref{eq:ribbon} (circles). The dashed line shows the bulk energy given by equation~\eqref{eq:bulk}}
\label{fig:ribbon}
\end{figure}

This trend can be rationalized using jellium ribbons. A jellium ribbon has a finite width $L_x$ and thickness $L_z$ but length $L_y$ that approaches infinity. The $r_s$ for a ribbon is defined as

\begin{equation}
\frac{L_x L_y L_z}{N}=\frac{4\pi r_s^3}{3},
\label{eq:ribbon_rs}
\end{equation}
where $N$ is the number of electrons in the unit cell and $L_y$ the length of the unit cell in the periodic direction. We show that DFT energies for ribbons with varying $L_x$, $L_z$ and $r_s$ are reasonably well described by a liquid drop model~\cite{perdew91}. This will imply a simple connection between edge and surface energies. The liquid drop model gives the total energy of the jellium ribbon as

\begin{equation}
E_j(r_s) = L_x L_y L_z u(r_s) + 2 (L_x L_y + L_y L_z) \sigma(r_s),
\end{equation}
where $u$ is the energy density of bulk jellium and $\sigma$ the surface energy. Using equation~\eqref{eq:ribbon_rs}, the energy per electron $\epsilon_j = E_j/N$ becomes

\begin{equation}
\epsilon_j = \epsilon_b + \frac{8\pi r_s^3}{3}\left(\frac{1}{L_x}+\frac{1}{L_z}\right)\sigma,
\label{eq:ribbon}
\end{equation}
where $\epsilon_b$ is the energy per electron for bulk jellium. We calculate ribbons with thickness of single fermi wave length $\lambda_f$ and change the width $L_x$ from $\lambda_f$ to $5\lambda_f$. Figure~\ref{fig:ribbon} shows the DFT results of fitting equation~\eqref{eq:ribbon} to changing ribbon widths $L_x$ for different values of $r_s$. The resulting bulk energies per electron (\mbox{Fig. \ref{fig:ribbon}}b) are close to the analytic expression in Rydberg units

\begin{equation}
\epsilon_b^{\mathrm{analytic}} = \frac{2.21}{r_s^2}-\frac{0.916}{r_s}-(0.115-0.0313\cdot\mathrm{ln}r_s)
\label{eq:bulk}
\end{equation}
and fitted surface energies (\mbox{Fig. \ref{fig:ribbon}}a) agree with previously reported values~\cite{lang70}. Therefore the energy of a ribbon is reasonably well described by the liquid drop model. Since the energy contribution from the jellium edges is $2 L_y L_z \sigma$, it is natural that the edge energies have similar $r_s$ trend as the surface energies. Nevertheless, the correspondence between edge and surface energies is somewhat surprising considering that the ribbons are only monolayer thick.

\subsection{2D and 3D vacancy formation energies correlate well}

\begin{figure}
\includegraphics[width = 8.6 cm]{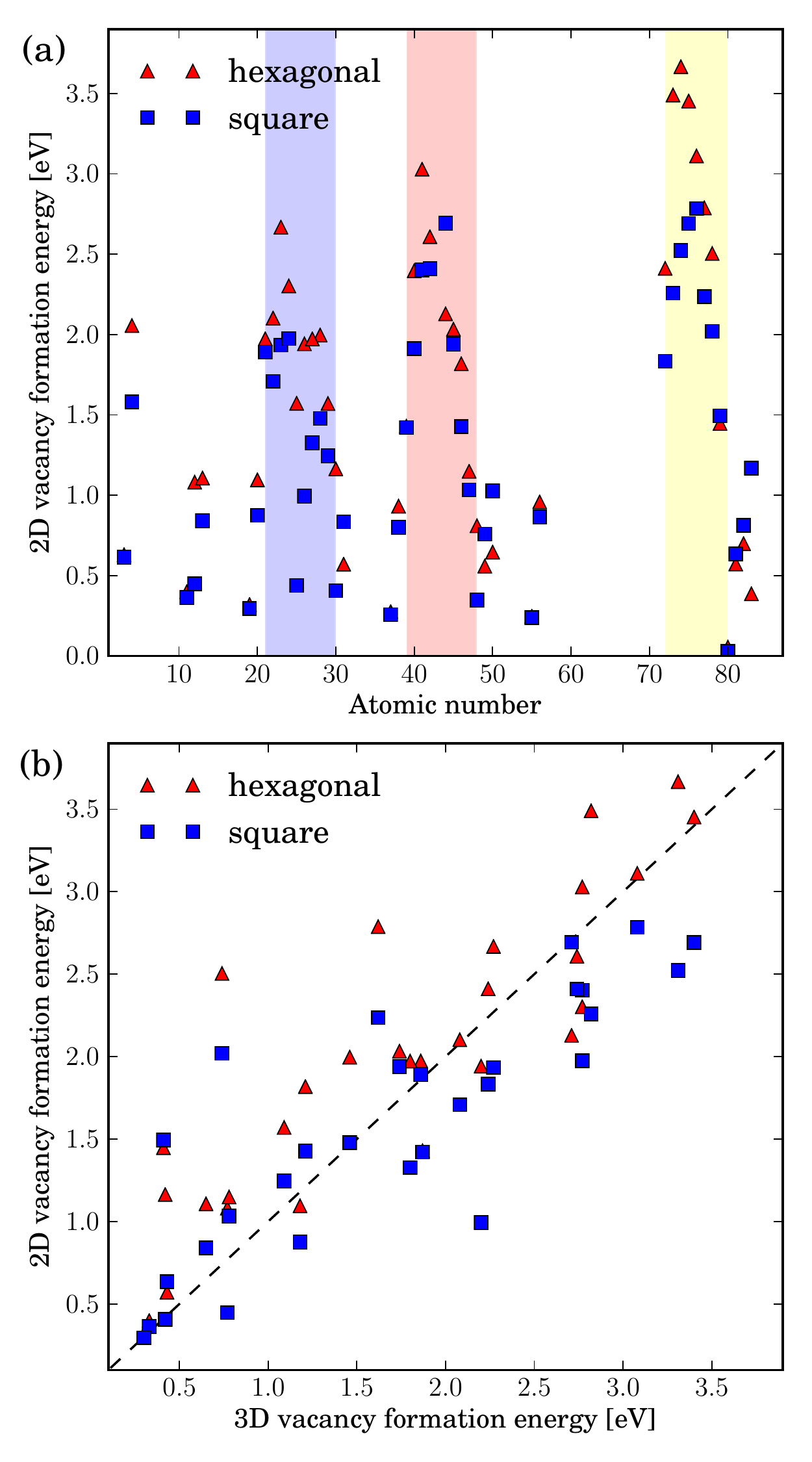}
\caption{(a) Calculated 2D vacancy formation energies for hexagonal and square ribbons as a function of atomic number. Shaded regions are as in Fig.~\ref{fig:E_edge}. (b) Comparison between 2D vacancy formation energies calculated here and 3D vacancy formation energies calculated in ref.~\cite{bharat15}. Dashed line shows equal energies.}
\label{fig:Ev}
\end{figure}

Next we study the vacancy formation energies. In practice they depend on the vacancy density, but this dependence can be removed using a method analogous to the one used to calculate the edge energy. We calculate five monolayers of different size, each with a single vacancy. The vacancy formation energy $\epsilon_{\mathrm{v}}$ is then obtained from a fit

\begin{equation}
E_v(N) = N\epsilon_{\mathrm{2D}}+\epsilon_{\mathrm{v}},
\label{eq:vacancy}
\end{equation}
where $E_v(N)$ is the energy of the monolayer with $N$ atoms and a single vacancy. As in the previous section, we can compare the $\epsilon_{\mathrm{2D}}$ from the fit to the value from periodic calculation to confirm convergence. The resulting vacancy formation energies range from nearly zero to almost 4 eV (\mbox{Fig. \ref{fig:Ev}}a). Comparison to calculated 3D bulk vacancy formation energies shows that for many metals the vacancy formation energies in 2D and 3D have similar values (\mbox{Fig. \ref{fig:Ev}}b). This supports our previous observation of 2D bonds being stronger than the 3D bonds~\cite{nevalaita18}.

\begin{figure}
\includegraphics[width = 8.6 cm]{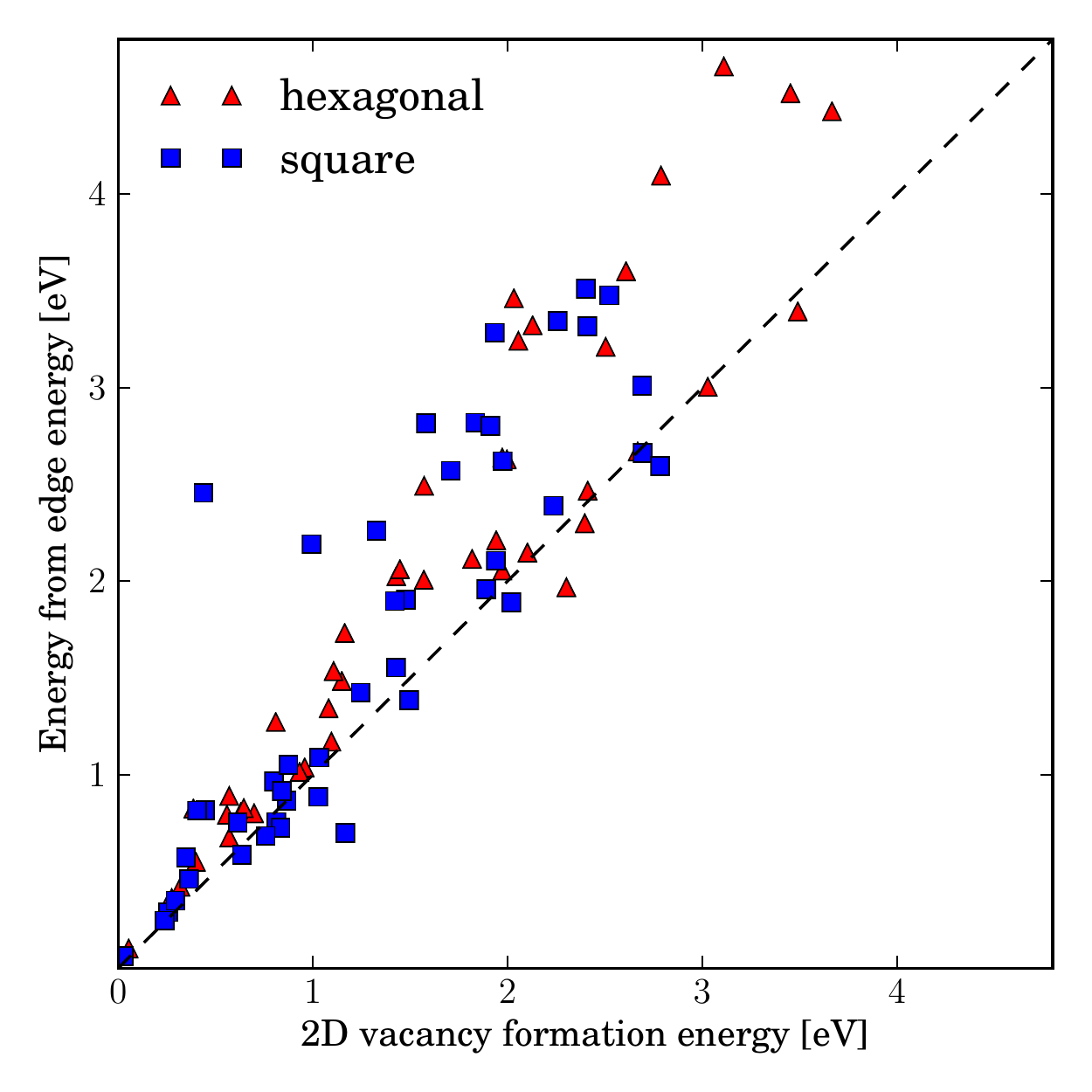}
\caption{Vacancy formation energies obtained by direct calculation compared to vacancy formation energies estimated using edge energy. Dashed line shows equal energy.}
\label{fig:Ev_vs_edge}
\end{figure}

Moreover, the vacancy formation and edge energies display qualitatively similar trends. The vacancy formation energies are generally high near the middle of d-series and again the highest for the metals near the center of 5d-series. This can be understood by considering the formation of a vacancy as a formation of finite-length edges to the monolayer. To confirm this interpretation, we plot the vacancy formation energy as a function of the edge energy times the length of the formed edge. We simply assume that the vacancy is circular with edge length $\pi d$, where $d$ is the nearest neighbor distance. While the vacancy formation energies from this simple approximation are generally overestimated they are fairly close to the directly calculated ones (\mbox{Fig. \ref{fig:Ev_vs_edge}}).

\subsection{2D and 3D work functions are nearly identical}

\begin{figure}
\includegraphics[width = 8.6 cm]{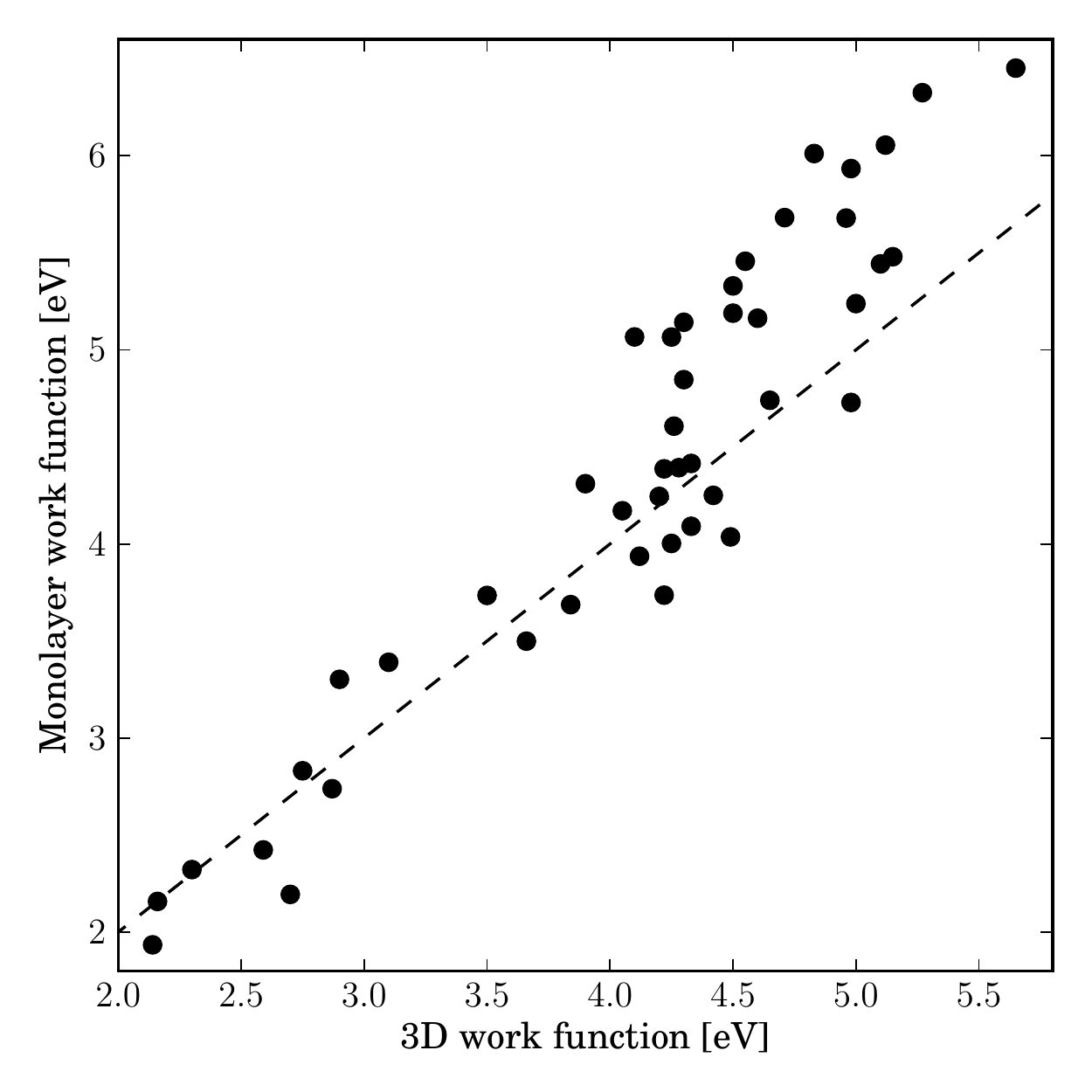}
\caption{Calculated work functions for hexagonal 2D lattices as a function of experimental work functions measured from polycrystalline samples~\cite{michaelson77}. Dashed line shows equal work functions.}
\label{fig:wf}
\end{figure}

Next, we consider the work functions of hexagonal 2D monolayers of all 45 metals. The work function is a global reactivity descriptor related to the cost of electron removal~\cite{harinipriya02,kahn16}. As a result, we find that the calculated work functions of 2D monolayers are close to the work functions measured for polycrystalline 3D bulk samples (\mbox{Fig. \ref{fig:wf}}). This is somewhat surprising since the polycrystalline samples contain crystals with different sizes and surface structures. For nanosized systems the variation in size is known to lead to oscillations in the work function. These changes are due to quantum size effects that arise when some sample dimensions are close to the electron wave lengths~\cite{paggel02,miller09,kim10}. While the quantum size effects are important for accurate calculations of small samples, the correlation between monolayer and polycrystalline work functions indicates that the quantum size effects do not drastically alter the work function. Therefore, the work function of a 3D system is a reasonable first approximation for the work function of a monolayer.

\subsection{2D polarizability increases with 3D bond length}

\begin{figure}
\includegraphics[width = 8.6 cm]{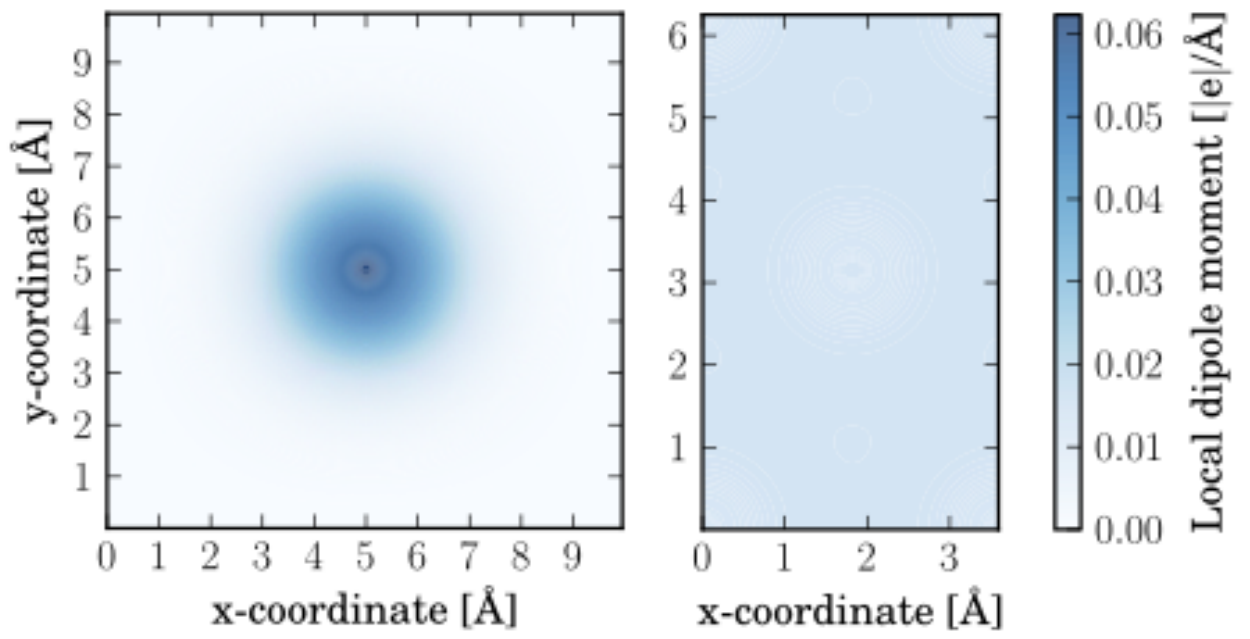}
\caption{Local dipole moment for Na atom (left panel) and hexagonal Na monolayer (right panel) at constant electric field of 0.5 V/\AA. For monolayer the local dipole moment is almost constant.}
\label{fig:ploc}
\end{figure}

\begin{figure}
\includegraphics[width = 8.6 cm]{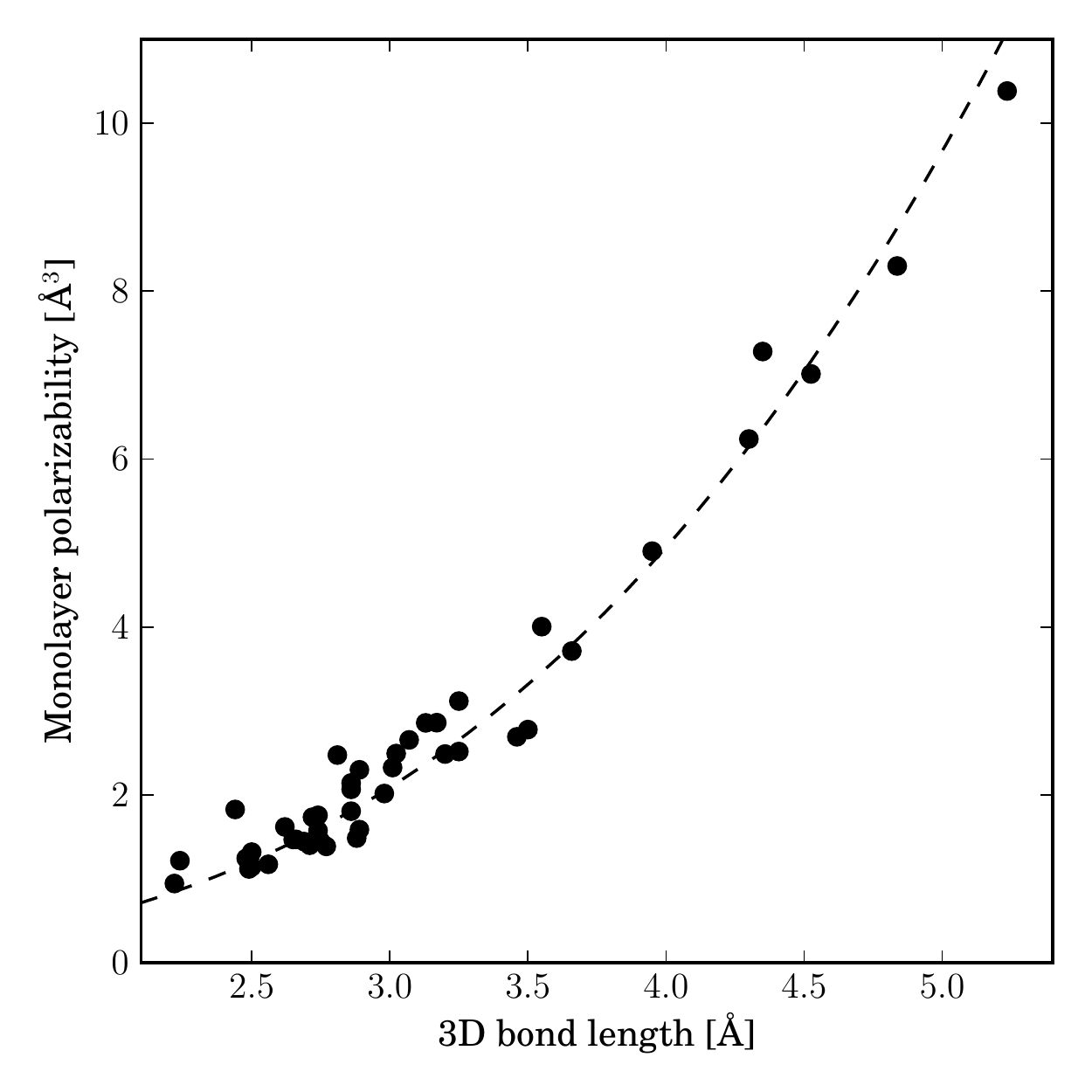}
\caption{Polarizability per atom for hexagonal 2D layers as a function of experimental 3D bond length~\cite{kittel}. Dashed line is the fit $\alpha = k d^3$, where $d$ is the bond length and $k = 0.077$ a fit parameter.}
\label{fig:alpha}
\end{figure}

To keep our approach as generic as possible, we also consider a perturbation of a constant electric field perpendicular to the atomic plane. We quantify the results in terms of the static polarizabilities calculated for hexagonal 2D metals only. Calculation of polarizabilities is motivated because the polarizability of a molecule is related to its reactivity. For example, larger polarizability of a molecule correlates with stronger physisorption~\cite{wang12}. Similarly, larger surface polarizabilities increase physisorption energies. We calculated the polarizability $\alpha$ by applying an electric field perpendicular to the atomic plane and varying the field strength $\mathcal{E}$ from 0.05 to 0.5 V/\AA. The polarizability was then given by a fit

\begin{equation}
p = \alpha \mathcal{E},
\label{eq:alpha}
\end{equation}
where $p$ is the dipole moment. The resulting polarizabilities per atom for the monolayers range from $1$ to $10$ \AA$^3$, which are significantly lower than the corresponding polarizabilities of free atoms that range from $5$ to $60$ \AA$^3$~\cite{hati94} (\mbox{Fig. \ref{fig:alpha}}). In order to visualize the difference between free atom and a monolayer, we define the local dipole moment

\begin{equation}
p_{\mathrm{loc}}(x,y)=\int\mathrm{d}z n(x,y,z)z,
\end{equation}
where $n$ is the electron density and $z$ is the direction perpendicular to the atomic plane. While free Na atom has large dipole moment near the nucleus, $p_{\mathrm{loc}}$ is almost constant for a hexagonal Na monolayer (\mbox{Fig. \ref{fig:ploc}}). 

To understand the difference between free and bound atoms, we consider a simple model. The static polarizability per atom is given by the $\mathcal{E}\rightarrow 0$ limit of the equation~\cite{pansini16}

\begin{equation}
\alpha = -\frac{\partial^2\epsilon_{\mathrm{ML}}(\mathcal{E})}{\partial^2 \mathcal{E}},
\end{equation}
where $\epsilon_{\mathrm{ML}}$ is the energy per atom for a monolayer. For a periodic system the energy per atom in terms of cohesive energy $\epsilon_{\mathrm{coh}}$ is 

\begin{equation}
\epsilon_{\mathrm{ML}}=\epsilon_{\mathrm{free}}-\epsilon_{\mathrm{coh}},
\end{equation}
where $\epsilon_{\mathrm{free}}$ is the energy of a free atom. Therefore, if $\epsilon_{\mathrm{coh}}(\mathcal{E})$ and $\epsilon_{\mathrm{free}}(\mathcal{E})$ are known, the polarizability of an extended system can be calculated. In practice, exact expressions for $\epsilon_{\mathrm{coh}}(\mathcal{E})$ and $\epsilon_{\mathrm{free}}(\mathcal{E})$ are not known. However, by approximating $\epsilon_{\mathrm{coh}}(\mathcal{E})$ the polarizability of many-atom system $\alpha_{\mathrm{ML}}$ can be calculated in terms of polarizability of free atom $\alpha_{\mathrm{free}}$. A simple approximation for $\epsilon_{\mathrm{coh}}(\mathcal{E})$ can be obtained by considering interaction between aligned dipoles. Assume that the application of an electric field gives rise to an equal dipole moment for each atom. The dipole-dipole interaction energy is $\epsilon_{\mathrm{dip}}(r) = p^2/r^3$, where $p$ is the dipole moment and $r$ the distance between dipoles. Since $p=\alpha_{\mathrm{ML}} \mathcal{E}$, the energy as a function of the applied field is $\epsilon_{\mathrm{dip}}(r,\mathcal{E})=\alpha_{ML}^2\mathcal{E}^2/r^3$. Approximating the cohesion energy as a sum over all dipoles gives

\begin{equation}
\epsilon_{\mathrm{coh}}(\mathcal{E}) = \sum_{(n,m)\neq(0,0)}\frac{\alpha_{\mathrm{ML}}^2 \mathcal{E}^2}{|n\vec{a}+m\vec{b}|^3},
\end{equation}
where the summation is over $(n,m)\in \mathbb{Z}^2\setminus (0,0) $ and $\vec{a}$ and $\vec{b}$ are the lattice vectors. Taking the second derivative with respect to $\mathcal{E}$ gives

\begin{equation}
\alpha_{\mathrm{ML}} = \alpha_{\mathrm{free}}-2S\frac{\alpha_{\mathrm{ML}}^2}{d^3},
\end{equation}
where $d$ is the bond length and $S$ is the lattice sum

\begin{equation}
S = \sum_{(n,m)\neq(0,0)}\frac{1}{(n^2+nm+m^2)^{3/2}},
\end{equation}
as discussed in ref~\cite{zucker17}. Solving for $\alpha_{\mathrm{ML}}$ gives

\begin{equation}
\alpha_{\mathrm{ML}} = \frac{d^3}{4S}\left(\sqrt{1+8S\frac{\alpha_{\mathrm{free}}}{d^3}}-1\right).
\label{eq:alphas}
\end{equation}
According to equation~\eqref{eq:alphas}, the polarizability of an extended system is always smaller than that of a free atom because $\epsilon_{\mathrm{ML}}(\mathcal{E})$ varies slower than $\epsilon_{\mathrm{free}}(\mathcal{E})$ due to the cost from from dipole interactions. Also $\alpha_{\mathrm{ML}}\rightarrow \alpha_{\mathrm{free}}$ when $S\rightarrow 0$ or $d\rightarrow \infty$, as expected. If we approximate the polarizability of a free atom to be proportional to the size of the atom~\cite{ghosh02,wong03}, which in turn is proportional to the bond length, the polarizability of a monolayer becomes proportional to the cube of the bond length. This cubic dependence is indeed what we observe (Fig.~\ref{fig:alpha}).

\section{Summary and Conclusions}

We studied the properties of edges and defects in mono-atomic free-standing 2D structures composed of metal atoms by density-functional theory. We considered 45 metals in hexagonal and square geometries and calculated their edge and vacancy formation energies. To keep our results general, we removed ribbon size and vacancy density dependence by utilizing a linear ansatz with a correct asymptotic behavior. We rationalized the ansatz for edge energies with a simple model of particles in 1D box. The edge energies ranged from almost zero to $0.6$ eV/\AA \ and had the highest values near the middle of d-series. Further, they decreased almost monotonically with increasing Wigner-Seitz radius, especially for the simple metals. A similar trend has been observed for surface energies. We explained this connection and the dependence between edge and surface energies by jellium ribbons and the liquid drop model.

The 2D vacancy formation energies ranged from nearly zero to almost $4$ eV and were highest near the middle of the d-series. For many metals the 2D vacancy formation energies were unexpectedly close to the 3D vacancy formation energies. This was in line with our earlier observation of 2D bonds being stronger than 3D ones. The vacancy formation energies were approximated well by the edge energies after considering the formation of a vacancy as the formation of a hole with a round edge. This connection implies that the edge energies can be used to estimate the formation energies also for vacancies of multiple atoms. Further, the energies of flat clusters can be quickly estimated using the edge energies.

Last, we perturbed the hexagonal monolayers by a constant electric field and calculated the dipole polarizabilities. We found that the polarizability per atom is significantly lower for monolayers compared to free atoms. We gave a simple model based on dipole interactions and obtained a cubic dependence between bond length and monolayer polarizability. The polarizability is relevant for physisorption of the monolayer on substrate and for adsorbates on the monolayer since larger polarizabilities lead to stronger physisorption. The work functions of hexagonal monolayers were found to be close to work functions measured for polycrystalline samples, indicating that the changes in surfaces structures and quantum size effects do not drastically alter them when going from 3D to 2D. The work function is a global reactivity descriptor that is important in charge transfer processes. These results, which have been collected to a single table (Table~\ref{tab:appendix}) for readers' benefit, contribute to the growing field of 2D metals and especially to advancing from idealized semi-infinite systems towards more realistic finite systems with defects and interactions between the environment.

\section*{Acknowledgments}

We acknowledge the Academy of Finland for funding (project 297115).

\section*{Appendix}
Table~\ref{tab:appendix} shows edge and vacancy formation energies for hexagonal and square lattices together with work functions and polarizabilities of hexagonal structures. 
\begin{table*}
\caption{Edge and vacancy formation energies ($\epsilon_{\mathrm{edge}}$ and $E_v$) for hexagonal (hex) and square (sq) lattices. Work function $\phi$ and polarizability $\alpha$ are tabulated for hexagonal lattice only.}
\label{tab:appendix}
\begin{tabular}{l|cccccc}

& $\epsilon_{\mathrm{edge}}^{\mathrm{hex}}$ (eV/\AA)& $\epsilon_{\mathrm{edge}}^{\mathrm{sq}}$ (eV/\AA)& $E_v^{\mathrm{hex}}$ (eV) & $E_v^{\mathrm{sq}}$ (eV) & $\phi$ (eV) & $\alpha$ (\AA$^3$)\\ 
\hline
Ag &	0.17 &	0.13 &	1.15 &	1.03 &	4.61 &	1.59 \\
Al &	0.18 &	0.11 &	1.11 &	0.84 &	4.39 &	1.81 \\
Au &	0.24 &	0.16 &	1.45 &	1.49 &	5.44 &	1.49 \\
Ba &	0.07 &	0.07 &	0.96 &	0.87 &	2.19 &	7.28 \\
Be &	0.48 &	0.44 &	2.06 &	1.58 &	4.73 &	0.95 \\
Bi &	0.08 &	0.07 &	0.39 &	1.17 &	4.39 &	2.66 \\
Ca &	0.1 &	0.09 &	1.1 &	0.87 &	2.74 &	4.9 \\
Cd &	0.14 &	0.06 &	0.81 &	0.35 &	3.74 &	2.02 \\
Co &	0.36 &	0.32 &	1.97 &	1.33 &	5.24 &	1.14 \\
Cr &	0.23 &	0.35 &	2.3 &	1.97 &	5.33 &	1.32 \\
Cs &	0.02 &	0.02 &	0.25 &	0.24 &	1.93 &	10.38 \\
Cu &	0.26 &	0.19 &	1.57 &	1.25 &	4.74 &	1.17 \\
Fe &	0.29 &	0.3 &	1.94 &	0.99 &	5.19 &	1.24 \\
Ga &	0.1 &	0.09 &	0.57 &	0.83 &	4.25 &	1.83 \\
Hf &	0.27 &	0.32 &	2.41 &	1.83 &	4.31 &	2.86 \\
Hg &	0.01 &	0.01 &	0.05 &	0.03 &	4.04 &	2.33 \\
In &	0.08 &	0.07 &	0.56 &	0.76 &	3.94 &	2.52 \\
Ir &	0.51 &	0.31 &	2.79 &	2.24 &	6.33 &	1.4 \\
K &	0.03 &	0.03 &	0.32 &	0.29 &	2.32 &	7.01 \\
Li &	0.08 &	0.08 &	0.63 &	0.61 &	3.3 &	2.49 \\
Mg &	0.14 &	0.09 &	1.08 &	0.45 &	3.5 &	2.49 \\
Mn &	0.31 &	0.34 &	1.57 &	0.44 &	5.07 &	1.22 \\
Mo &	0.44 &	0.43 &	2.61 &	2.41 &	5.16 &	1.74 \\
Na &	0.05 &	0.04 &	0.4 &	0.36 &	2.83 &	3.71 \\
Nb &	0.36 &	0.44 &	3.03 &	2.4 &	4.85 &	2.07 \\
Ni &	0.35 &	0.27 &	2.0 &	1.48 &	5.48 &	1.12 \\
Os &	0.58 &	0.34 &	3.11 &	2.78 &	6.01 &	1.46 \\
Pb &	0.08 &	0.08 &	0.7 &	0.81 &	4.0 &	2.78 \\
Pd &	0.26 &	0.19 &	1.82 &	1.43 &	6.05 &	1.45 \\
Pt &	0.39 &	0.24 &	2.5 &	2.02 &	6.45 &	1.39 \\
Rb &	0.02 &	0.02 &	0.27 &	0.26 &	2.16 &	8.3 \\
Re &	0.56 &	0.39 &	3.45 &	2.69 &	5.68 &	1.57 \\
Rh &	0.43 &	0.27 &	2.03 &	1.94 &	5.93 &	1.44 \\
Ru &	0.42 &	0.36 &	2.13 &	2.69 &	5.68 &	1.47 \\
Sc &	0.21 &	0.21 &	1.97 &	1.89 &	3.73 &	3.12 \\
Sn &	0.08 &	0.1 &	0.65 &	1.03 &	4.25 &	2.48 \\
Sr &	0.08 &	0.08 &	0.93 &	0.8 &	2.42 &	6.24 \\
Ta &	0.4 &	0.41 &	3.49 &	2.26 &	5.07 &	2.14 \\
Ti &	0.26 &	0.32 &	2.1 &	1.71 &	4.41 &	2.3 \\
Tl &	0.06 &	0.06 &	0.57 &	0.64 &	3.69 &	2.69 \\
V &	0.35 &	0.45 &	2.67 &	1.93 &	5.14 &	1.62 \\
W &	0.53 &	0.44 &	3.67 &	2.52 &	5.46 &	1.76 \\
Y &	0.19 &	0.19 &	1.43 &	1.42 &	3.39 &	4.0 \\
Zn &	0.22 &	0.1 &	1.16 &	0.41 &	4.09 &	1.47 \\
Zr &	0.25 &	0.32 &	2.4 &	1.91 &	4.17 &	2.86

\end{tabular}
\end{table*}

\end{document}